\documentclass[a4paper]{jpconf}

\usepackage{amsmath}
\usepackage{amssymb}
\usepackage{graphicx,color}

\usepackage{wrapfig}

\begin{document}

\title{The evolution of non-linear disturbances in magnetohydrodynamic flows}

\author{Alexander~V. Proskurin}

\address{Altai State Technical University, 656038, Russian Federation, Barnaul, Lenin prospect,46}

\ead{k210@list.ru}

\author{Anatoly~M. Sagalakov}

\address{Altai State University, 656049, Russian Federation, Barnaul, Lenin prospect, 61}

\begin{abstract}
In this article the stability loss of the Hartmann flow are investigated by applying the equations for disturbances. The velocity and electric potential quasi-static MHD model is used. The equations allow us to calculate time-dependent disturbance fields using a base flow and an initial disturbance. Two type of initial perturbations are considered: the eigenfunction of the linearized MHD equations and a fluid injection into the flow. These two approaches lead to identical stability results. However, we found a significant difference in the practical implementation of the two approaches. Dealing with the eigenproblem  of the linearized MHD system is a laborious task. In terms of calculation costs it is equal to a series of nonlinear perturbation simulations, and if the Hartmann number is increased, the proportion becomes worse. The non-linear stability analysis produced by these two methods shows that the injection technique can also be used in numerical analysis, and that this method is less expensive in terms of calculation costs.
\end{abstract}

\section{Introduction}

Liquid metal flows occur in many technological applications including metallurgy, materials processing, and nuclear and thermonuclear reactors cooled by liquid metals. Such facilities contain pipes and supporting devices such as pumps and valves in which laminar or turbulent flows can occur. For the design and operation of such liquid metal installations it is important to know exactly how flow regimes will be carried out from the flow parameters: pressure gradient, viscosity, electrical conductivity, and external magnetic field.

It is generally accepted that turbulent flow appears due to the non-linear development of perturbations. There are two theoretical approaches here \cite{hof2003scaling}. In one, the turbulent motion develops from infinitely small perturbations of the laminar flow. These tiny perturbations grow to have a relatively large amplitude in which a non-linear interaction becomes large. According to the second point of view, the turbulent state at large Reynolds numbers attracts the flow which is routed from the laminar state by any small finite amplitude perturbation. 

\begin{figure}
\begin{center}
\includegraphics[width=0.8\textwidth]{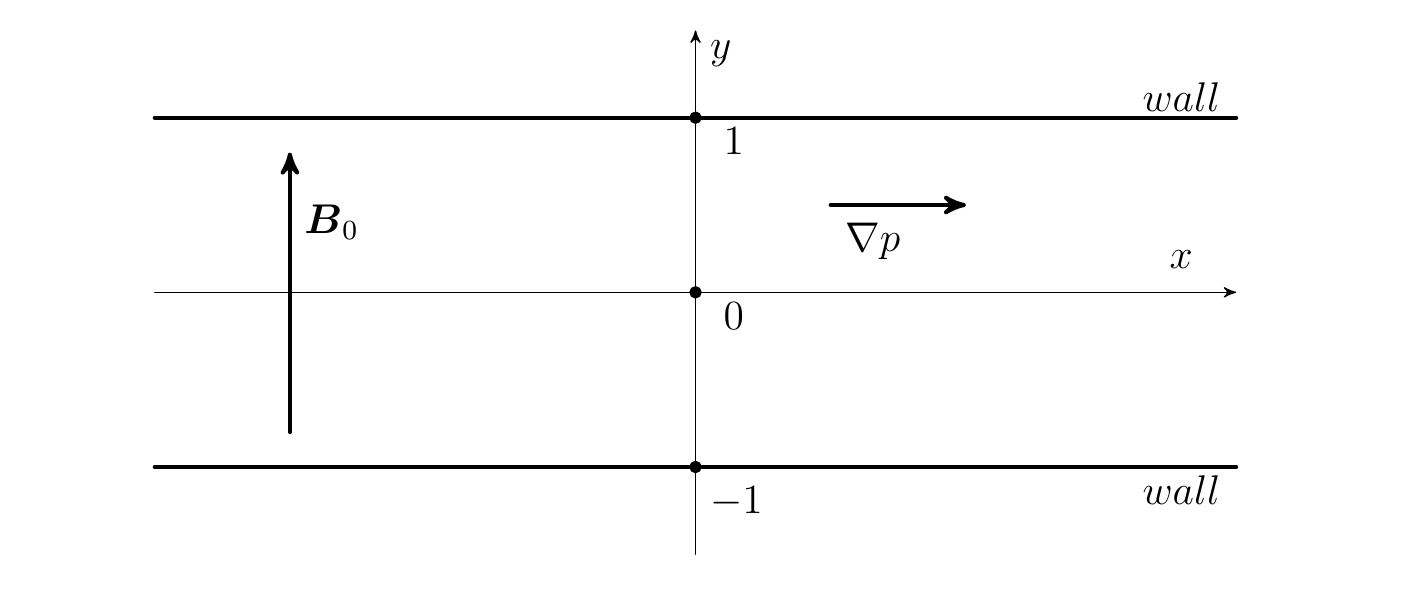}
\end{center}
\caption{The plane channel}\label{a24.PlaneChannel}
\end{figure}

We decide to compare these two approaches. The Hartmann flow in a plane channel is considered as a test example of magnetohydrodynamic flow. A sketch of the flow is shown in Figure \ref{a24.PlaneChannel}. Two parallel infinite planes are installed at $y = \pm 1$. The liquid between the planes flows under a constant pressure gradient $\nabla p$ in a $x$ direction. A  uniform magnetic field $\boldsymbol{B}_0$ is perpendicular to the planes.
We suppose that $Re_m \ll Re$. In this case a magnetic field generated by the fluid movement does not affect the imposed magnetic field. This is correct for most engineering applications \cite{lee2001magnetohydrodynamic}.  It is now possible to write the Navier-Stokes equation in the form
\begin{equation}
\label{article24.NS_Syst_MagForce}
\begin{aligned}
\frac{\partial \boldsymbol{v}}{\partial{t}}+\left( \boldsymbol{v} \nabla \right)\boldsymbol{v} &= -\frac{1}{\rho}\nabla p + \nu \Delta \boldsymbol{v} + \boldsymbol{F}(\boldsymbol{v},\boldsymbol{B}_0),\\
div \boldsymbol{v} &= 0,
\end{aligned}
\end{equation}
where $\boldsymbol{v}$ is the velocity, $p$ is the pressure, $\nu$ is the viscosity, $\rho$ is the density, $\boldsymbol{F}$ is the magnetic force, and $\boldsymbol{B}_0$ is the imposed magnetic field.

Ohm's law is:
\begin{equation}
\label{article24.j_eq}
\boldsymbol{j} = \sigma\left( -\nabla\varphi+\boldsymbol{v}\times\boldsymbol{B}_0  \right),
\end{equation}
where $\boldsymbol{j}$ is the density of the electric current, $\varphi$ is the electric potential, and $\sigma$ is the conductivity. Using the law of conservation of electric charge ($div \boldsymbol{j} = 0$), it is possible to derive the equation for the electric potential as:
\begin{equation}
\label{article24.ElPot_eq}
\Delta \varphi = \nabla(\boldsymbol{v}\times\boldsymbol{B}_0).
\end{equation}

The system (\ref{article24.NS_Syst_MagForce}) can be written in the form: 
\begin{equation}
\label{article24.WeakMHD_Syst}
\begin{aligned}
\frac{\partial \boldsymbol{v}}{\partial{t}}+\left( \boldsymbol{v} \nabla \right)\boldsymbol{v} &= -\nabla p + \frac{1}{Re} \Delta \boldsymbol{v} + St \left( -\nabla\varphi+\boldsymbol{v}\times\boldsymbol{B}_0  \right)\times\boldsymbol{B}_0 ,\\
&div \boldsymbol{v} = 0,\\
&\Delta \varphi = \nabla(\boldsymbol{v}\times\boldsymbol{B}_0),
\end{aligned}
\end{equation}
where $Re = \frac{L_0V_0}{\nu}$ is the Reynolds number, $St=\frac{\sigma B_0^2 L_0}{\rho V_0}$ is the magnetic interaction parameter (the Stuart number), and $L_0$, $V_0$, and $B_0$ represent the scales of length, velocity and magnetic field, respectively. The system (\ref{article24.WeakMHD_Syst}) is also known as the MHD system in a quasi-static approximation in electric potential form. This system is widely used in theoretical investigations and accurately approximate many cases of liquid metal flows (for example, see the appropriate discussion and reference in \cite{krasnov2011comparative}).

\begin{figure}[b]
\centering
\includegraphics[width=0.6\textwidth]{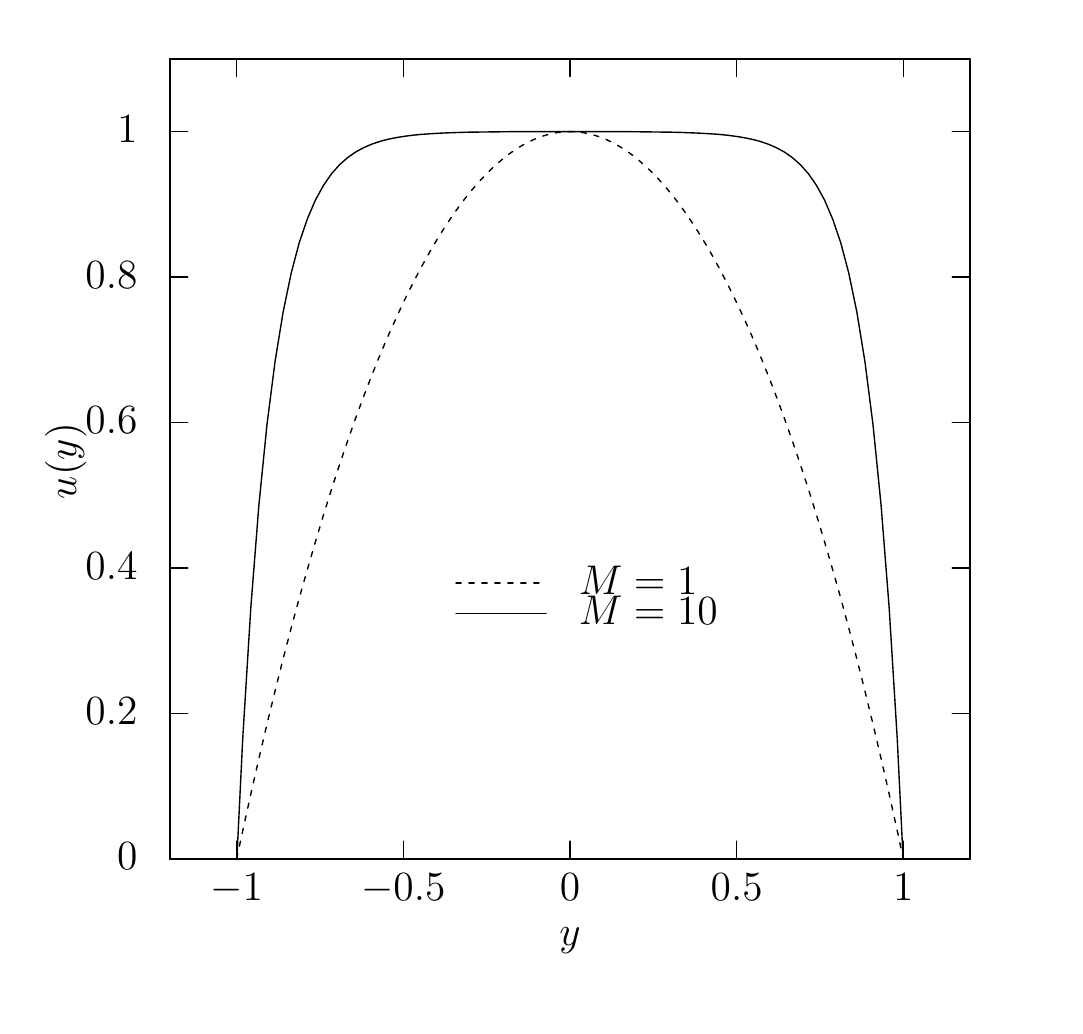}
\caption{The Hartmann flow graph at $M=1,\,10$}\label{a24.VelocityHartmann}
\end{figure}

According to \cite{davidson2016introduction} the state solution of (\ref{article24.WeakMHD_Syst}) is 
\begin{equation}
\label{article24.hartman_flow_1}
\frac{u(y)}{u(0)}=\frac{\cosh{\left (M \right )} - \cosh{\left (M y \right )}}{\cosh{\left (M \right )} - 1}
\end{equation}
where $M = \sqrt{St \cdot Re}$ is the Hartmann number, and $u(0)$ is the centreline velocity. A velocity graph at $M=1$ and $M=10$ is shown in Figure \ref{a24.VelocityHartmann}.

\section{Nonlinear disturbances and numerical method}

The flow variables can be decomposed to a form
\begin{equation}
\label{article24.nonlinear_form}
\begin{aligned}
\boldsymbol{v} &= \boldsymbol{U}+\boldsymbol{v},\\
\varphi &= \varphi_0+\varphi,\\
p &= p_0 +p,
\end{aligned}
\end{equation}
where $\boldsymbol{U}$, $\varphi_0$, and $p_0$ are values of a steady-state solution and $\boldsymbol{v}$, $\varphi$ and $p$ are disturbances. Now the system (\ref{article24.WeakMHD_Syst}) is:
\begin{equation}
\label{article24.disturbNS}
\begin{aligned}
\frac{\partial \boldsymbol{v}}{\partial{t}}+\left( \boldsymbol{U} \nabla \right)\boldsymbol{v}+\left( \boldsymbol{v} \nabla \right)\boldsymbol{U} + \left( \boldsymbol{v} \nabla \right)\boldsymbol{v} &= -\nabla p + \frac{1}{Re} \Delta \boldsymbol{v} + St \left( -\nabla\varphi+\boldsymbol{v}\times\boldsymbol{B}_0  \right)\times\boldsymbol{B}_0 ,\\
&div \boldsymbol{v} = 0,\\
&\Delta \varphi = \nabla(\boldsymbol{v}\times\boldsymbol{B}_0).
\end{aligned}
\end{equation}

We investigate the non-linear growth of initial disturbances by time integration of the equations (\ref{article24.disturbNS}). The initial disturbance can be considered as an eigenfunction of the linearized MHD equations as described in \cite{orszag1980transition}. The linearized Navier-Stokes equations are the subject of hydrodynamic stability theory \cite{schmid2012stability}. These equations  contain a small parameter $\frac{1}{Re}$ that causes exponentially growing and high frequency oscillating solutions, which are hard to find numerically. Also, these difficulties are significant for the investigation of the non-linear perturbations in the transient and turbulent states of flow.

To avoid numerical difficulties, we should use high-order methods \cite{orszag:1971,theofilis:2011:highorder, theofilis:2011:global}. High-order methods have good computational properties, fast convergence, small errors, and the most compact data representation. Our MHD solver \cite{proskurin2017spectral} has been developed on the basis of an open source spectral element framework \texttt{Nektar++} \cite{cantwell:2015, karniadakis:2013}. The incompressible Navier-Stokes solver (\texttt{IncNavierStokesSolver}) from this framework has been taken as the source for the MHD solver. Details of the numerical method, and it's reliability and convergence is discussed in \cite{proskurin2017spectral}.  

In this paper, we investigate perturbations separately from the main flow. Such calculations are more costly than for complete flow fields, since it is necessary to calculate advective terms three times ($\left( \boldsymbol{U} \nabla \right)\boldsymbol{v}$, $\left( \boldsymbol{v} \nabla \right)\boldsymbol{U}$ and $\left( \boldsymbol{v} \nabla \right)\boldsymbol{v}$), but it is a small piece in all volume of the calculations. We made a comparative analysis and did not find that the total computation time increased noticeably. However, this approach allows us to simplify data manipulation and operate more detailed calculations with regard to disturbances. For example, it is possible to investigate the disturbances associated with an unstable flow, which can be obtained using stabilizing techniques.

\begin{figure}[p]
\begin{center}
\includegraphics[width=\textwidth]{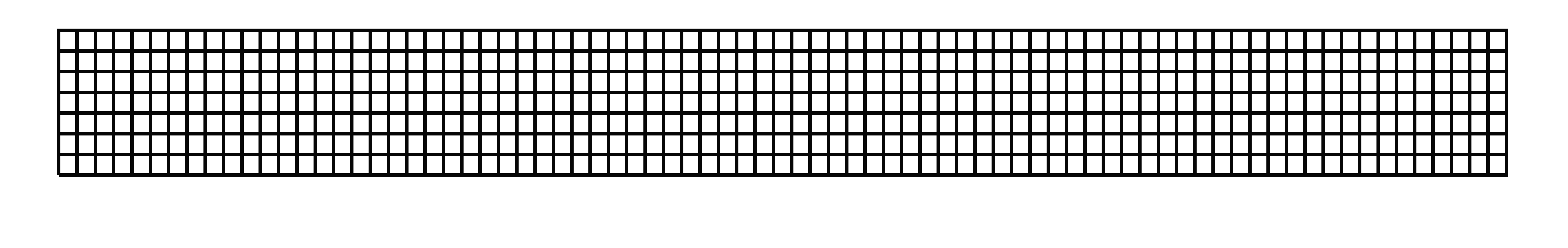}
\end{center}
\caption{The mesh}\label{article24.PlaneChanGrid}
\end{figure}

A mesh is shown in Figure \ref{article24.PlaneChanGrid}. We set up periodic boundary conditions at the inflow and the outflow and 
\begin{equation}\label{article24.boundaryConditions}
\boldsymbol{v}=0, \frac{\partial p}{\partial \boldsymbol{n}}=0
\end{equation}
on the walls. It is supposed that the flow is two-dimensional, and boundary conditions for $\varphi$ are not required.

\section{Growth of the eigenfunction disturbance}

Now, we investigate the evolution of the perturbations which ​​are given by the linearized problem. In this case, the linearized MHD equations are considered, and the eigenvalue problem is solved. Details are described in \cite{proskurin2017spectral}. Two cases are explored: $Re=10^4$, $M = 1$(case A) and $Re=10^5$, $M=10$(case B). In case A, the flow graph is close to a parabolic at $M=0$(without the magnetic field), but the critical Reynolds number is higher. In case B, the magnetic field strongly affects the flow. In both cases, the flow is stable in terms of infinitely small perturbations. The largest eigenvalues ​​are $(-0.0733, \pm 5.56)$ and $(-0.0461,\pm 3.21)$ for cases A and B, respectively. We take the eigenfunctions corresponding to these eigenvalues ​​as the initial conditions for the equations (\ref{article24.disturbNS}).

\begin{figure}[p]
\begin{tabular}{c}
\includegraphics[width=\textwidth]{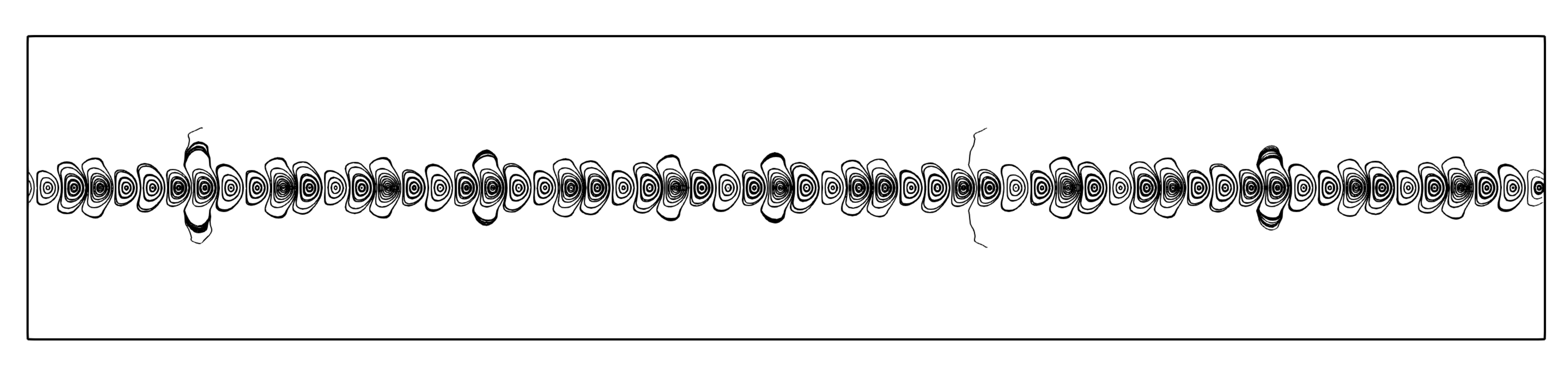}\\
(a)\\
\includegraphics[width=0.985\textwidth]{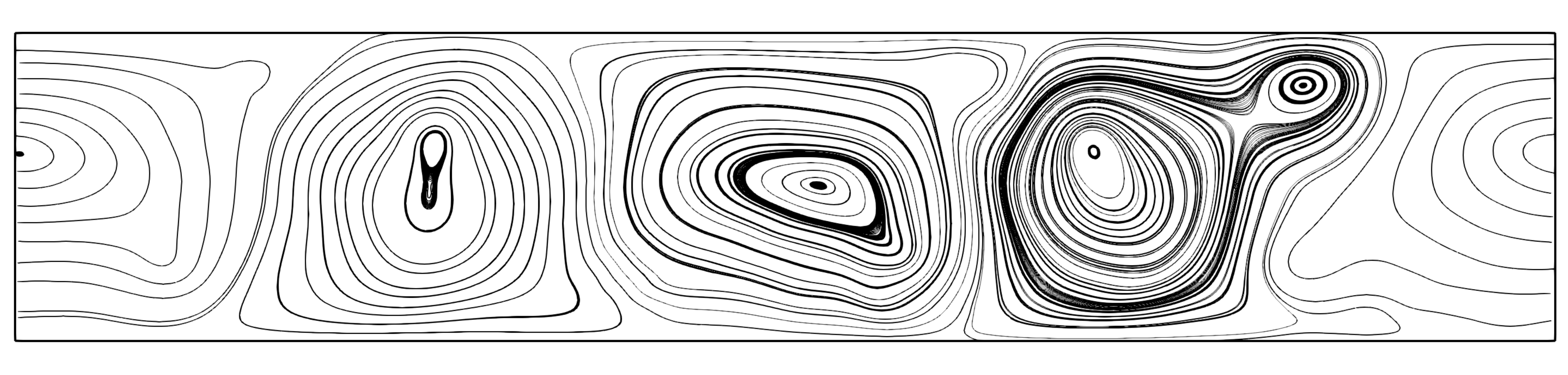}\\
(b)
\end{tabular}
\caption{The disturbance streamlines at $t=0$ (a) and $t=50$ (b), $Amp=1.0$, $Re=10^4$, $M=1$.}\label{a24.DisturbRe10000M1}
\end{figure}

\begin{figure}[p]
\hspace{0.05\textwidth}
\begin{minipage}{0.4\textwidth}
\includegraphics[width=\textwidth]{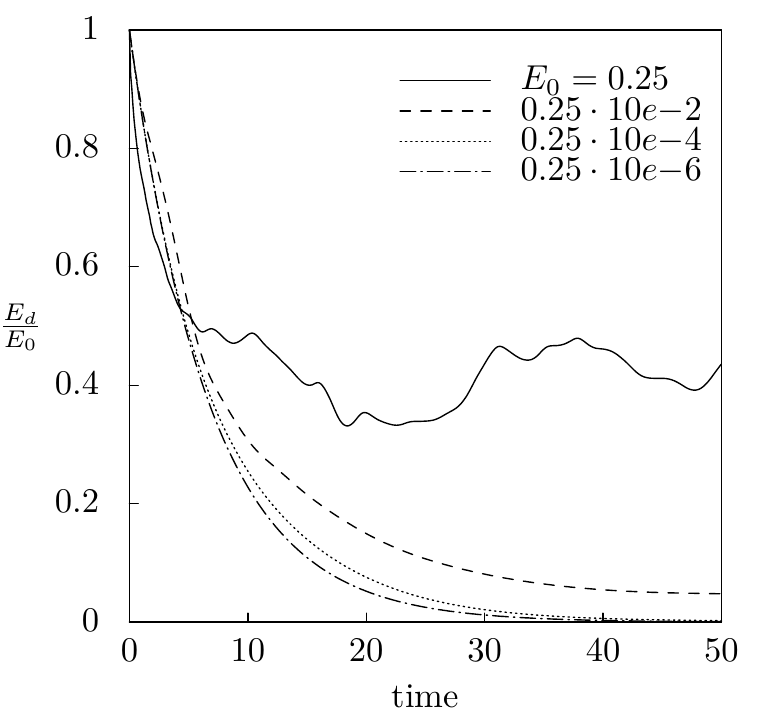}
\caption{Temporal energy evolution of the eigenfunction disturbance for different values of the initial energy $E_0$ at $Re=10^4$, $M=1$}\label{a24.growthRe10000M1}
\end{minipage}\hspace{0.1\textwidth}
\begin{minipage}{0.4\textwidth}
\includegraphics[width=\textwidth]{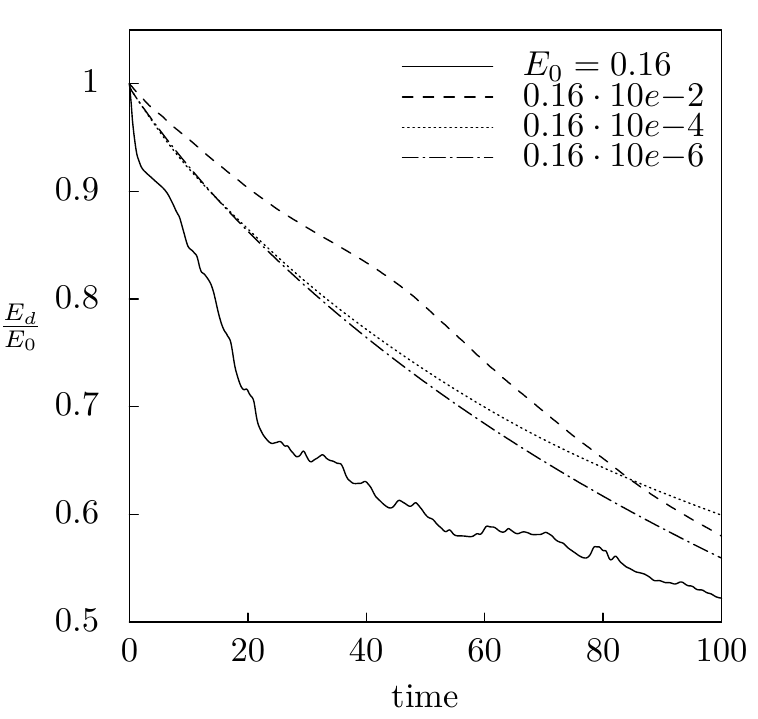}
\caption{Temporal energy evolution of the eigenfunction disturbance for different values of the initial energy $E_0$ at $Re=10^5$, $M=10$}\label{a24.growthRe100000M10}
\end{minipage}
\end{figure}

For the eigenfunctions we set the maximum velocity amplitudes $Amp=1.0,0.1,0.01,0.001$ and calculate the evolution of each perturbation in time. Figure \ref{a24.DisturbRe10000M1} shows streamlines for the disturbance in the case A for $Amp=1.0$, $t=0$ (a) and $t=50$ (b). Figures \ref{a24.growthRe10000M1} and \ref{a24.growthRe100000M10} show graphs of the temporal evolution of the energy of the perturbations for cases A, B and several values ​​of the initial energy $E_0$, corresponding to the values ​​of the amplitude $Amp$. $E_d$ is the energy of the disturbances. The energy graphs allow us to determine that in most cases the perturbations decay, except for one plot at $Re=10^4$, $M=1$, where it is possible to observe persisting flow energy oscillations.

\section{Growth of the injection disturbance}
\begin{figure}[p]
\begin{minipage}[t]{0.3\textwidth}
\includegraphics[width=\textwidth]{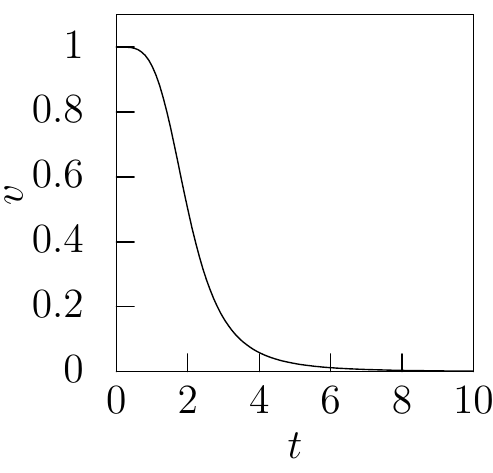}
\caption{Time decrease of the disturbance at point $x=-2$}\label{article24.injection_disturbance_time}
\end{minipage}\hspace{0.045\textwidth}
\begin{minipage}[t]{0.3\textwidth}
\includegraphics[width=\textwidth]{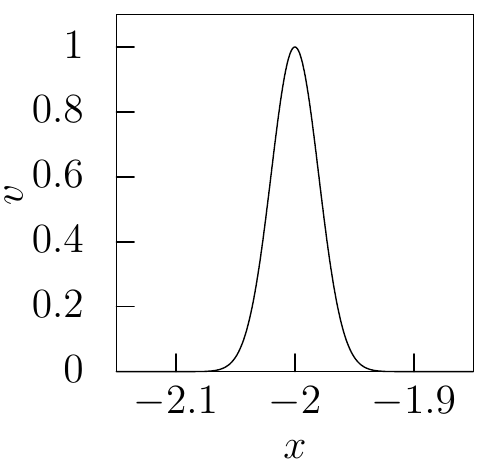}
\caption{Graph of the disturbance at $t=0$}\label{article24.injection_disturbance}
\end{minipage}\hspace{0.045\textwidth}
\begin{minipage}[t]{0.3\textwidth}
\includegraphics[width=\textwidth]{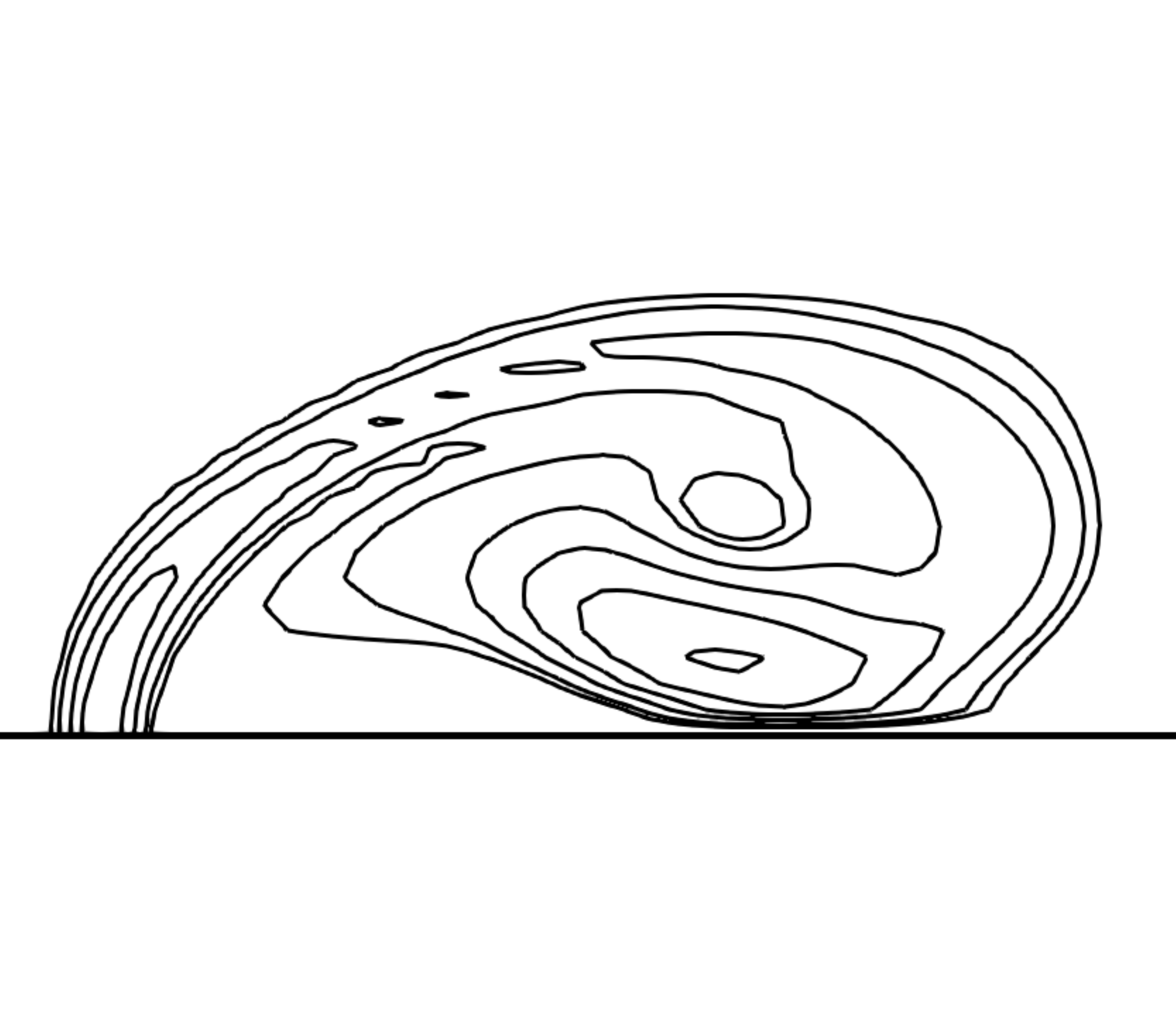}
\caption{Isocontures of the disturbance velocity amplitude at $Re=3000$, $M=10^{-3}$}\label{article24.injectionRe3000M0p001A1p0}
\end{minipage}
\end{figure}

\begin{figure}[p]
\hspace{0.05\textwidth}
\begin{minipage}[t]{0.4\textwidth}
\includegraphics[width=\textwidth]{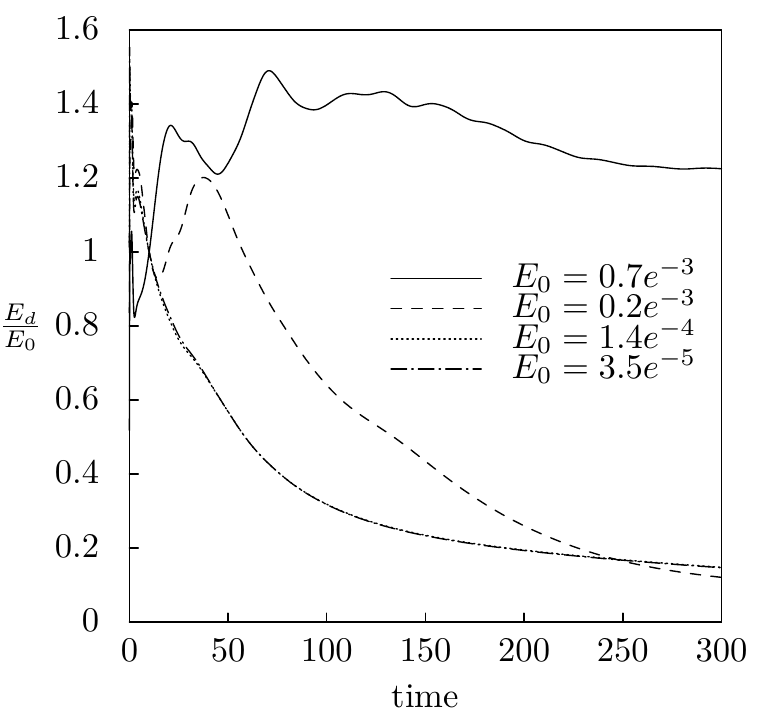}
\caption{Temporal energy evolution of the injection disturbance for different values of the initial energy $E_0$ at $Re=10^4$, $M=1$}\label{a24.injGrowRe10000M1}
\end{minipage}\hspace{0.1\textwidth}
\begin{minipage}[t]{0.4\textwidth}
\includegraphics[width=\textwidth]{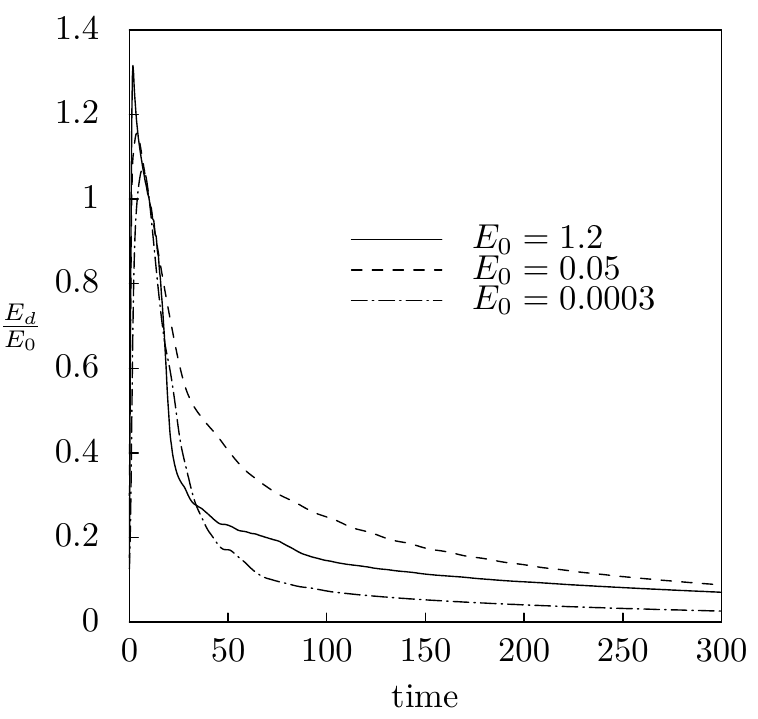}
\caption{Temporal energy evolution of the injection disturbance for different values of the initial energy $E_0$ at $Re=10^5$, $M=10$}\label{a24.injGrowRe100000M10}
\end{minipage}
\end{figure}

In this section the injection of fluid is considered. A perturbation on the bottom wall has the form
\begin{equation}\label{article24.disturbInj}
v_{inj} = \frac{1.0 e^{- 1250.0 \left(x + 2.0\right)^{2}}}{0.0625 t^{4} + 1}.
\end{equation}
This method of introducing perturbations to the flow is widely used in experiments. Figures \ref{article24.injection_disturbance_time}, \ref{article24.injection_disturbance} show graphs of the disturbance amplitude dependence from the $x$ coordinate and the time $t$. In Figure \ref{article24.injectionRe3000M0p001A1p0} the isocontures of velocity amplitude are presented at $Re=3000$, $M=10^{-3}$. The perturbation (\ref{article24.disturbInj}) is set at the top and the bottom so that the fluid volume is constant. Figures \ref{a24.injGrowRe10000M1} and \ref{a24.injGrowRe100000M10} show graphs of the temporal evolution of the perturbation energy for cases A ($Amp=1,\,0.7,\,0.2,\,0.1$) and B ($Amp=7,\,2,\,0.1$). The energy is scaled by the perturbation energy $E_0$, measured at $t=10$. The energy graphs allow us to determine that in most cases the perturbations decay, except fore one plot at $Re=10^4$, $M=1$, where it is possible to see the stability loss.

\begin{figure}[p]
\includegraphics[width=0.985\textwidth]{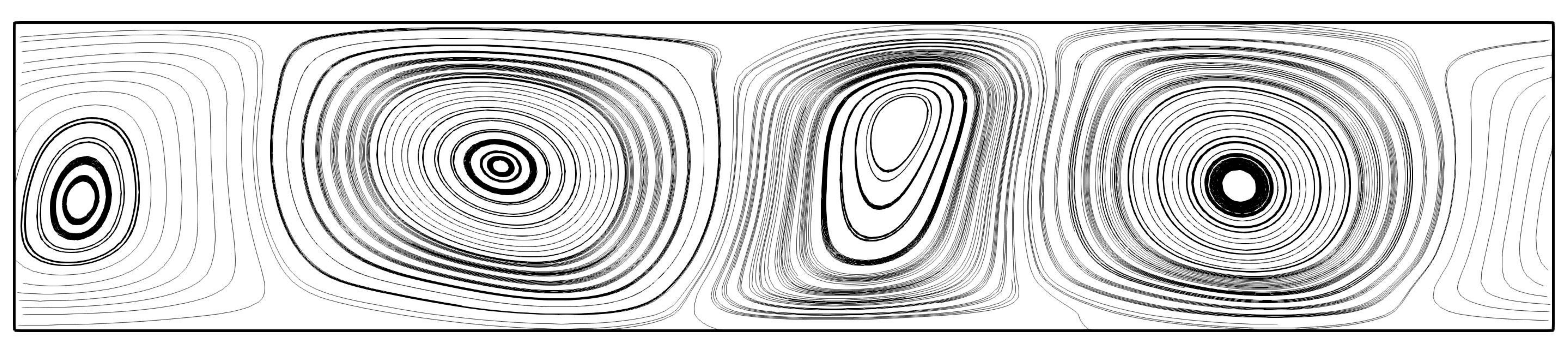}\\
\caption{The disturbance streamlines at $t=300$, $M=1$, $Re=10^4$}\label{a24.InjDisturbRe10000M1}
\end{figure}

\section{Conclusion}
In this article, we have examined the evolution of the finite amplitude disturbances in the Hartmann flow. Two types of disturbances have been investigated: the eigenfunctions of the linearized MHD equations, and the injection of fluid through the side wall. These theoretical approaches were compared by carrying out the analysis of two-dimensional perturbations. Two sets of parameters are considered: $Re=10^4$ and $M=1$(case A), $Re=10^5$ and $M=10$(case B). The two approaches led to the identical stability results. For case B, the flow is stable with regard to the finite amplitude perturbations. For case A, a time-dependent non-laminar flow can be observed if the perturbation amplitude is sufficiently large. We found a significant difference in the practical implementation of the approaches. Dealing with the eigenproblem  of the linearized MHD system is a laborious task. In terms of the calculation cost, it is equal to the series of non-linear perturbation simulations, while if the Hartmann number is increased, this proportion becomes worse. The non-linear stability analysis produced by these two methods shows that the injection technique can also be used in numerical analysis and this method is less expensive in terms of calculation cost. The comparison of these approaches for three-dimensional perturbations, complex geometry, and high Hartmann numbers is an attractive line for future research.

\bibliographystyle{iopart-num}
\bibliography{reference24}

\providecommand{\newblock}{}
\begin{thebibliography}{10}
\expandafter\ifx\csname url\endcsname\relax
  \def\url#1{{\tt #1}}\fi
\expandafter\ifx\csname urlprefix\endcsname\relax\def\urlprefix{URL }\fi
\providecommand{\eprint}[2][]{\url{#2}}

\bibitem{hof2003scaling}
Hof B, Juel A and Mullin T 2003 {\em Physical review letters\/} {\bf 91} 244502

\bibitem{lee2001magnetohydrodynamic}
Lee D and Choi H 2001 {\em Journal of Fluid Mechanics\/} {\bf 439} 367--394

\bibitem{krasnov2011comparative}
Krasnov D, Zikanov O and Boeck T 2011 {\em Computers \& fluids\/} {\bf 50}
  46--59

\bibitem{davidson2016introduction}
Davidson P~A 2016 {\em Introduction to magnetohydrodynamics\/} vol~55
  (Cambridge university press)

\bibitem{orszag1980transition}
Orszag S~A and Kells L~C 1980 {\em Journal of Fluid Mechanics\/} {\bf 96}
  159--205

\bibitem{schmid2012stability}
Schmid P~J and Henningson D~S 2012 {\em Stability and transition in shear
  flows\/} vol 142 (Springer Science \& Business Media)

\bibitem{orszag:1971}
Orszag S~A 1971 {\em Journal of Fluid Mechanics\/} {\bf 50} 689--703

\bibitem{theofilis:2011:highorder}
Gonzalez L, Theofilis V and Sherwin S~J 2011 {\em International journal for
  numerical methods in fluids\/} {\bf 65} 923--952

\bibitem{theofilis:2011:global}
Theofilis V 2011 {\em Annual Review of Fluid Mechanics\/} {\bf 43} 319--352

\bibitem{proskurin2017spectral}
Proskurin A~V and Sagalakov A~M 2018 {\em Magnetohydrodynamics\/} {\bf 54}
  361--372

\bibitem{cantwell:2015}
Cantwell C~D, Moxey D, Comerford A, Bolis A, Rocco G, Mengaldo G, De~Grazia D,
  Yakovlev S, Lombard J~E, Ekelschot D, Jordi B, Xu H, Mohamied Y, Eskilsson C,
  Nelson B, Vos P, Biotto C, Kirby R~M and Sherwin S~J 2015 {\em Computer
  Physics Communications\/} {\bf 192} 205--219 ISSN 0010-4655

\bibitem{karniadakis:2013}
Karniadakis G and Sherwin S 2005 {\em Spectral/hp Element Methods for
  Computational Fluid Dynamics: Second Edition\/} Numerical Mathematics and
  Scientific Computation (OUP Oxford) ISBN 9780199671366

\end{thebibliography}
\end{document}